\documentclass[rmp,superscriptaddress,twocolumn,preprint]{revtex4-1}

\usepackage{amsmath, amsthm, amscd, amssymb}
\usepackage{graphicx, braket}
\usepackage{graphics}
\usepackage{bm}
\usepackage{bbm}\usepackage{natbib}
\usepackage{setspace}
\begin{document}

\title{Stochastic Methods and Dynamical Wave-function Collapse}

\author{\large \bf Angelo Bassi}
\email{bassi@ts.infn.it}
\affiliation{Department of Physics \ University of Trieste\ Strada Costiera 11 \ 34151 Trieste Italy} \affiliation{Istituto Nazionale di Fisica Nucleare \ Trieste Section \ Via Valerio 2\ 34127 Trieste Italy}

\author{\large \bf Kinjalk Lochan}
\email{kinjalk@tifr.res.in} \affiliation{Tata Institute of Fundamental Research\  Homi Bhabha Road\ Mumbai 400005\ India}

\author{\large \bf Seema Satin}
\email{satin@imsc.res.in}
\affiliation{Inst. of Math. Sciences\ IV Cross Road\ CIT Campus\ Taramani\ Chennai 600 113
\ India}

\author{\large \bf Tejinder P. Singh}
\email{tpsingh@tifr.res.in}
\affiliation{Tata Institute of Fundamental Research\ Homi Bhabha Road\  Mumbai 400005\ India}

\author{\large \bf Hendrik Ulbricht}
\email{h.ulbricht@soton.ac.uk}
\affiliation{School of Physics and Astronomy\ University of Southampton\ SO17 1BJ\ UK}

\date{\today}

\maketitle

\centerline{\bf Abstract}

\noindent This brief article reviews stochastic processes as relevant to dynamical models of wave-function collapse, and is supplemental material for the review article arXiv:1204.4325

\bigskip
\bigskip

\newcommand{\be}{\begin{equation}}
\newcommand{\ee}{\end{equation}}
\newtheorem{definition}{Definition}
\newcommand{\R}{\mathbf{R}}

The best known example of a stochastic process is Brownian motion : random motion of small particles suspended in a liquid, under the influence of a viscous drag, and a fluctuating force resulting from  collisions with the molecules of this liquid. The quantitative explanations of Brownian motion by Einstein and Smoluchowski were simplified by Langevin, through his use of what is now called the Langevin equation. This dynamical equation for the randomly evolving position of the suspended particle is equivalent to the
Fokker-Planck equation for the time evolution of the probability distribution of the random variable. The Langevin equation can be put on a firm footing as a stochastic differential equation using It\^o's differential calculus for stochastic variables. The brief review below introduces stochastic processes, and should be of some assistance in understanding the review article on dynamical wave-function collapse ~\cite{RMP:2012}
and should be read in conjunction with that review. For a detailed survey the reader is referred to the books by ~\cite{Gardiner:83, Risken:96, Arnold:71, Gikhman:72, Grimmett:01} and the review article by ~\cite{Chandra:43}.

\subsection{Some Basic Concepts from Probability Theory}
Let $\Omega$  denote a sample space, i.e. a set of all possible events in an
 experiment.  Then one can define the following structures on $\Omega$

\begin{definition}
A collection $\mathcal{F}$ of subsets of $\Omega$ is called a $\sigma$- field
if
\begin{enumerate}
\item $\phi \in \mathcal{F}$.
\item if $A_1,A_2 \dots \in \mathcal{F}$ then $\cup_{i=1}^{\infty} A_i \in
\mathcal{F}$.
\item if $A \in \mathcal{F}$ then $A^C \in \mathcal{F}$ .
\end{enumerate}
\end{definition}
 The above properties also imply that closed countable intersections are also
included in $\mathcal{F}$.

Examples:

a) $\mathcal{F} = \{\phi, \Omega \}$ is the smallest $\sigma-$ field.

b) If $A \subset \Omega$ then $\mathcal{F} = \{\phi, A, A^c, \Omega \}$ is
a $\sigma$-field.
\begin{definition}
A probability measure $P$ on $(\Omega,\mathcal{F})$ is a function $P: \mathcal{F
} \rightarrow [0,1]$ satisfying:
\begin{enumerate}
\item $P(\phi) = 0, P(\Omega) = 1$.
\item if $A_1,A_2 \dots$ is a collection of disjoint members of $\mathcal{F}$
in that $A_i \cap A_j = \phi$ for all points $i,j$ satisfying $i \neq j$ then
\[ P(\cup_{i=1}^{\infty} A_i) = \sum_{i=1}^\infty P(A_i) \]
\end{enumerate}
The triple $(\Omega, \mathcal{F},P)$ is defined to be the probability space.
\end{definition}
\subsection{Random Variables}
\begin{definition}
A random variable is a function  $X:\Omega \rightarrow \R$ with the property
that $\{ \omega \in \Omega : X(\omega) \leq x \} \in \mathcal{F}$ for each
$x \in \R$, such functions are called $\mathcal{F}- $ measurable.

\end{definition}
\subsubsection{Distribution of random variables}
\begin{itemize}
\item $X$ is called a discrete random variable if it takes values in some
countable subset $\{x_1,x_2,\dots,\}$ only of $\R$. Then $X$ has probability
mass function $f : \R \rightarrow [0,1]$ defined by $f(X) = P(X=x)$.

\item $X$ is called a continuous random variable if its distribution
function $F:\R \rightarrow [0,1]$ given by $F(x) = P(X \leq x)$ can be
expressed as
\[ F(x) = \int_{-\infty}^{x} f(u) du  \mbox{  }  x \in \R \]
for some integrable function $f:\R \rightarrow [0,\infty)$, where $f$ is the
probability density function.

\item Joint distribution Function : A random variable $\mathbf{X} = (X_1,X_2,
\dots X_n)$ on $(\Omega, \mathcal{F},P)$ is a function $F_{\mathbf{X}}: \R^n
\rightarrow [0,1]$ given by $F_{\mathbf{X}} (\mathbf{x}) = P(\mathbf{X} \leq
\mathbf{x})$ for $\mathbf{x} \in \R^n $.
\end{itemize}
\subsubsection{Time Dependent Random Variables}
Let $\xi \equiv \xi(t)$ be a time dependent random variable. Assume an ensemble
of systems, such that each system leads to a number $\xi$ which depends on time.

The outcome for one system cannot be precisely predicted, but ensemble averages exist and can be calculated. For fixed $t=t_1$ we define the probability
density by
\be
W_1(x_1,t) = \langle\delta(x_1-\xi(t_1))\rangle
\ee
where the angular brackets denote the ensemble average. The probability to find
the random variable $\xi(t_1)$ in the interval $ x_1 \leq \xi(t_1) \leq x_1 +
dx_1$ is given by $W_1(x_1,t) dx_1 , \dots, $ in interval $x_n \leq \xi(t_n)
\leq x_n + dx_n $ is given by $W_n(x_n,t_n;\dots;x_1,t_1)$ $dx_n dx_{(n-1)}
 \dots dx_1$,
where
\begin{eqnarray}
& &W_n(x_n,t_n; \dots ; x_1,t_1) =\cr
& & \langle\delta(x_1 -\xi(t_1)) \dots \delta(x_n -
\xi(t_n)) \rangle 
\end{eqnarray}
Given $W_n$ for all $n$, for every $t_i$ in the interval $t_0 \leq t_i \leq t_0
+T$ the time dependence of the process described by $\xi(t)$ in the interval
$[t_0,t_0+T]$ can be known completely.

One can obtain probability densities with lower number of variables from those
of higher number of variables by integrating as follows. For $i < r$
\begin{eqnarray}
\nonumber
& & W_i(x_1,x_2,\dots, x_i) =  \\ & &\int \dots \int W_r(x_1,x_2,\dots,x_i,x_{i+1},\dots
,x_r) \nonumber \\ & & dx_{i+1} \dots dx_r
\end{eqnarray}

 Conditional Probability density is  defined as:
\[
P(x_1|x_2 \dots x_r) = \frac{W_r(x_1 \dots x_r)}{W_{r-1}(x_2,\dots x_r)}
\]
where $P(x_1|x_2 \dots x_r)$ denotes the probability density of $x_1$ given the
$x_2, \dots, x_r$.
\subsubsection{Stationary Processes}
If the probability densities do not change by replacing  $t_i$ by $t_i +T$
($T$ arbitrary) the process is called a \textit{stationary process}. In such a
case $W_1$ does not depend on $t$ and $W_2$ can depend on the time difference
$t_2 - t_1$.

\textbf{The Wiener-Khinchin Theorem :}
According to this theorem, the spectral density is the Fourier Transform of the correlation function for
stationary processes.

Instead of the random variable $\xi(t)$ one may consider its Fourier Transform
\[\tilde{\xi}(\omega) = \int_{-\infty}^\infty \exp(-i\omega t) \xi(t) dt \]
For a stationary process $\langle\xi(t) \xi^*(t')\rangle$ is a function only of the
difference $t-t'$ i.e.
\[\langle\xi(t)\xi^*(t')\rangle = \langle\xi(t-t')\xi^*(0)\rangle \]
Introducing the new variables
\[ \tau = t-t' \]
\[ t_0 = (t+t')/2 \]
we have
\begin{eqnarray}
& & \langle\tilde{\xi}(\omega)\tilde{\xi}^*(\omega ')\rangle = \int_{-\infty}^{\infty}
\exp(-i(\omega - \omega ')t_0) dt_0 \nonumber \\ & & \int_{-\infty}^\infty \exp(- i(\omega '
+ \omega)\tau/2) \langle\xi(\tau)\xi^*(0)\rangle d \tau
\end{eqnarray}
which gives

\[
\langle \tilde{\xi}(\omega)\tilde{\xi}^*(\omega ')\rangle = \pi \delta(\omega - \omega')
S(\omega)
\]
where
\[ S(\omega) = 2 \int_{-\infty}^\infty \exp( i\omega \tau)
\langle\xi(\tau)\xi^*(0)\rangle d \tau \]
is the spectral density.

\subsection{Stochastic Processes and their Classification}
Stochastic processes are systems which evolve probabilistically, i.e. systems in which a time-dependent random variable exists.

The stochastic processes described by a random variable $\xi$ can be classified as follows

\textbf{Purely Random Process:} If $P_n (n \geq 2) $ does not depend on
the values $x_i = \xi(t_i), (i < n)$ at $t_i < t_n $, then,
\[
P(x_n,t_n|x_{n-1},t_{n-1};\dots;x_1,t_1) = P(x_n,t_n)
\]
thus
\[
 W_n(x_n,t_n ;\dots;x_1,t_1) = P(x_n,t_n) \dots P(x_1,t_1)
\]
Hence, complete information of the process is  contained in $P(x_1,t_1) =
W_1(x_1,t_1)$. A purely random process cannot describe physical systems where
the random variable is a continuous function of time.

\textbf{Markov Process:} If the conditional probability density depends only on
 the value $\xi(t_{n-1}) = x_{n-1}$ at $t_{n-1}$ , but not on $\xi(t_{n-2})
= x_{n-2}$ at $t_{n-2}$ and so on, then such a process is known as a Markov
Process. This is given by
\begin{eqnarray}
& & P(x_n,t_n|x_{n-1},t_{n-1};\dots;x_1,t_1) = \nonumber \\ & & P(x_n,t_n|x_{n-1},t_{n-1})
\end{eqnarray}
Then it follows that
\begin{eqnarray}
& & W_n(x_n,t_n; \dots ;x_1,t_1) =\nonumber \\ & & P(x_n,t_n|x_{n-1},t_{n-1})P(x_{n-1},t_{n-1}|
x_{n-2},t_{n-2})\nonumber \\
& &  \dots P(x_2,t_2|x_1,t_1)W_1(x_1,t_1)
\end{eqnarray}
For $n=2$
\[
P(x_2,t_2|x_1,t_1) = \frac{W_2(x_2,t_2;x_1.t_1)}{W_1(x_1,t_1)}
\]
For a Markov process the complete information of the process is contained in
$W_2(x_2,t_2;x_1,t_1)$.

\textbf{General processes:} There can be other processes such that the
complete information is contained in $W_3$, $W_4$ etc. However this
classification is not suitable to describe non-Markovian processes. For
describing non-Markovian processes, more random variables (other than
$\xi(t) = \xi_1(t)$), $\xi_2(t),\dots,\xi_r(t))$ can be considered. Then by proper
choice of these additional variables, one can have a Markov process for $r$
random variables.

\subsection{Markov Processes}
The Markov ssumption is formulated in terms of conditional probabilities as
follows :
\begin{eqnarray}
& & P (x_1,t_1;x_2,t_2; \dots |y_1,\tau_1;y_2,\tau_2; \dots)  \nonumber \\ & =& P(x_1,t_1;x_2,t_2
; \dots|y_1,\tau_1)
\end{eqnarray}
 From the defintion of conditional probability
\begin{eqnarray}
& & P(x_1,t_1;x_2,t_2|y_1,\tau_1) =\nonumber\\ & & P(x_1,t_1|x_2,t_2;y_1,\tau_1)P(x_2,t_2|y_1
,\tau_1)\nonumber\\
\end{eqnarray}
Using Markov property one can write
\begin{eqnarray}
& & P(x_1,t_1;x_2,t_2;x_3,t_3;\dots ;x_n,t_n) =\nonumber \\ & & P(x_1,t_1|x_2,t_2)P(x_2,t_2|
x_3,t_3)P(x_3,t_3|x_4,t_4) \nonumber \\
& & \dots P(x_{n-1},t_{n-1}|x_n,t_n)P(x_n,t_n)
\end{eqnarray}
provided $t_1 \geq t_2\geq t_3 \geq t_4 \geq \dots t_{n-1} \geq t_n$.

\subsubsection{The Chapmann Kolmogorov Condition}
The following condition holds for all stochastic processes :
\begin{eqnarray}
& &P(x_1,t_1) = \int dx_2 P(x_1,t_1;x_2,t_2) \nonumber \\
& = & \int dx_2 P(x_1,t_1|x_2,t_2) P(x_2,t_2)
\end{eqnarray}
Now
\begin{eqnarray*}
P(x_1,t_1|x_3,t_3)& =& \int dx_2 P(x_1,t_1;x_2,t_2|x_3,t_3)\\
& & \int dx_2 P(x_1,t_1|x_2,t_2;x_3,t_3)\\
& & P(x_2,t_2|x_3,t_3)
\end{eqnarray*}
Introducing the Markov assumption, if $t_1 \geq t_2 \geq t_3$, then we can
drop dependence on $x_3,t_3$. Thus,
\begin{eqnarray}
& & P(x_1,t_1|x_3,t_3) = \nonumber \\ & & \int dx_2 P(x_1,t_1|x_2,t_2) P(x_2,t_2|x_3,t_3)
\end{eqnarray}
This is the Chapmann Kolmogorov Equation. The differential form of this equation plays an important role in the description of stochastic processes that follows.
\subsubsection{The Fokker-Planck equation}
From the definition of transition probability one can write
\be \label{eq:chapman}
W(x,t+\tau) = \int P(x,t+\tau|x',t) W(x',t)dx'
\ee
This connects $W(x,t+\tau)$ with $W(x',t)$. To obtain a differential
equation for the above, the following procedure can be carried out.
Assume that all moments $M_n(x,t,\tau) , n \geq 1$ are known
\begin{eqnarray*}
M_n(x',t,\tau)& =& <[\xi(t+\tau) -\xi(t)]^n> |_{\xi(t) = x'}\\
& = &\int (x - x)^n P(x,t+\tau|x',t) dx
\end{eqnarray*}
Introducing $\Delta = x -x'$ in equation (\ref{eq:chapman}), the integrand can
 be expanded in a Taylor series, which after integration over $\delta$ can
 be put in the following form:
\begin{eqnarray} \label{eq:expansion}
& & W(x,t+\tau) - W(x,t) =    \frac{\partial W(x,t)}{\partial t} \tau + O(\tau^2) \nonumber \\  & & =\sum_{n=1}^{\infty} \left(- \frac{\partial}{\partial x}\right)^n \left( \frac{M_n(x,t,\tau
)}{n!}\right) W(x,t)
\end{eqnarray}
$M_n$ can be expanded into Taylor series with respect to $\tau$ ~\cite{Risken:96}
$(n \geq 1)$
\[M_n(x,t,\tau)/ n! = D^{(n)}(x,t) \tau + O(\tau^2)\]
Considering only linear terms in $\tau$, Eqn. (\ref{eq:expansion}) can be
written as
\begin{eqnarray} \label{eq:kramers}
\frac{\partial W(x,t)}{\partial t} & = & \sum_{n=1}^\infty \left(-\frac{\partial}
{\partial x}\right)^n D^{(n)}(x,t)W(x,t) \nonumber \\
& =& L_{KM}W(x,t) \nonumber
\end{eqnarray}
\be
L_{KM} = \sum_{n=1}^\infty \left(-\frac{\partial}{\partial x}\right)^n D^{(n)}(x,t)
\ee
The above is the Kramers-Moyal Expansion.
Pawula Theorem states that for a positive transition probability $P$ the
Kramers-Moyal expansion stops after the second term; if not then it must
 contain an infinite number of terms.

If the K-M expansion stops after the second term, then it is called the
Fokker -Planck equation, given by
\be
\frac{\partial W(x,t)}{\partial t} = -\frac{\partial}{\partial x} D^{(1)}(x,t)
+ \frac{\partial^2}{\partial x^2} D^{(2)}(x,t)
\label{Fokker-Planck}
\ee
$D^{(1)}$ is called the drift coefficient, and $D^{(2)}$ is called the diffusion coefficient. The transition probability $P(x,t|x',t')$ is the distribution $W(x,t)$ for
 the initial condition $W(x,t')= \delta(x-x')$. Therefore the transition
probability must also satistfy (\ref{eq:kramers}). Hence,
\be
\frac{\partial P(x,t|x',t')}{\partial t} = L_{KM}(x,t)P(x,t|x',t')
\ee
where the initial condition is given by $P(x,t|x',t) = \delta(x-x')$.

The Fokker-Planck equation can also be written as
\be
\frac{\partial W}{\partial t} + \frac{\partial S}{\partial x} =0
\ee
where
\[ S(x,t) = [D^{(1)}(x,t) - \frac{\partial}{\partial x}D^{(2)}(x,t)]W(x,t).
\]
Here $S$ can be interpreted as a probability current.
We will discuss the one variable Fokker-Planck equation with time-independent
drift and diffusion coefficients given by
\be \label{eq:timeind}
\frac{\partial W(x,t)}{\partial t} = -\frac{\partial}{\partial x} D^{(1)}(x)
+ \frac{\partial^2}{\partial x^2} D^{(2)}(x)
\ee

For stationary processes the probability current $S=constant$. Methods of
solution of such a F-P equation for stationary processes are discussed in
~\cite{Risken:96}. Non-stationary solutions of the F-P equation are in general
 difficult to obtain. A general expression for non-stationary solution can be
found only for special drift and diffusion coefficients.

The Fokker-Planck equation can be taken as a starting point for introducing the concept of a stochastic differential equation. If the random variable $\xi(t)$ satisfies the initial condition
\begin{equation}
W(\xi(t), y) = \delta (\xi - y)
\end{equation}
that is, it is sharply peaked at the value $y$, it can be shown by solving the Fokker-Planck equation that a short time $\Delta t$ later, the solution is still sharply peaked, and is a Gaussian with mean
$y + D^{(1)} \Delta t$ and variance $D^{(2)}$. The picture is that of a system moving with a systematic drift velocity $D^{(1)}$, and on this motion is superimposed a Gaussian fluctuation with variance $D^{(2)}$. Thus,
\begin{equation}
y(t + \Delta t ) = y(t) + D^{(1)} \Delta t + \eta (t) \Delta t ^{1/2}
\end{equation}
where $\langle \eta \rangle = 0$ and $\langle \eta^{2} \rangle = D^{(2)}$.   This picture gives sample paths which are always continuous but nowhere differentiable. As we will see shortly, this heuristic picture can be made much more precise and leads to the concept of the stochastic differential equation.

\subsubsection{Wiener Process}
A process which is described by Eqn. (\ref{eq:timeind}) with $D^{(1)} = 0$
and $D^{(2)}(x) = D =$ constant, is called a Wiener process. Then the
 equation for transition probability $P = P(x,t|x',t')$ is  the diffusion
equation, given by
\be
\frac{\partial P}{\partial t} = D \frac{\partial^2 P}{\partial x^2}
\ee
with the initial condition $P(x,t'|x',t') = \delta(x-x')$. Then the solution
for $t>t'$ is given by the gaussian distribution
\begin{eqnarray}
& & P(x,t|x',t') = \nonumber\\ & & \frac{1}{\sqrt{4\pi D(t-t')}} \exp\left(-\frac{(x-x')^2}{4D(t-t')}\right)
\end{eqnarray}
Thus the general solution for probaility density with initial distribution
$W(x',t')$ is given by
\[W(x,t) = \int P(x,t|x',t') W(x',t') dx'\]
An initially sharp distribution spreads in time.
The one-variable Wiener process is often simply called Brownian motion, since it obeys the same differential equation of motion as Brownian motion.
\subsubsection{Ornstein-Uhlenbeck Process}
This process is described by Eqn. (\ref{eq:timeind}) when the drift coefficient is linear and diffusion coefficient is constant, i.e.
\[ D^{(1)}(x) = -\gamma x ; D^{(2)}(x) = D = \mbox{ const } \]
The Fokker-Planck equation then can be written as
\be \label{eq:Ornst}
\frac{\partial P}{\partial t} = \gamma \frac{\partial}{\partial x} (xP) +
D \frac{\partial^2}{\partial x^2} P
\ee
with initial condition $P(x,t'|x',t') = \delta(x-x')$.
 The above equation can be solved by taking a Fourier transform w.r.t. $x$ i.e.
\[P(x,t|x',t') = (2 \pi)^{-1} \int e^{ikx} \tilde{P}(k,t|x',t')dk \]
Thus it results in the following equation
\[\frac{\partial \tilde{P}}{\partial t} = -\gamma k \frac{\partial \tilde{P}}
{\partial k} - D k^2 \tilde{P}\]
with initial condition $\tilde{P}(k,t'|x't') = e^{ikx'}$ for $t > t'$. Then
one gets
\begin{eqnarray}
& & \tilde{P}(k,t|x',t') = \nonumber \\ & & \exp[-ikx'e^{-\gamma(t-t')}- Dk^2(1-e^{-2\gamma(t-t')
})/(2 \gamma)] \nonumber  \\
\end{eqnarray}
By applying the inverse Fourier transform one finally obtains the solution of the
 Fokker Planck equation describing the Ornstein Uhlenbeck process
\begin{eqnarray}
\label{eq:sol}
& & P(x,t|x',t') = \sqrt{\frac{\gamma}{2 \pi D(1- e^{-2 \gamma(t-t')})} }\times \nonumber \\ & & \exp\left[-
\frac{\gamma(x-e^{-\gamma(t-t')} x')^2}{2D(1- e^{-2\gamma(t-t')})}\right] \nonumber \\
\end{eqnarray}
In the limit $\gamma \rightarrow 0$ we get the Gaussian distribution for a
Wiener Process.

Eqn. (\ref{eq:sol}) is valid for positive and negative values of $\gamma$.
For positive $\gamma$ and large time difference $\gamma(t-t')\gg 1$  the
equation passes over to the stationary distribution given by
\begin{equation}
W_{st} = \sqrt{\frac{\gamma}{2 \pi D}} \exp\left[-\frac{\gamma x^2}{2D}\right]
\end{equation}
For $\gamma \leq 0$ no stationary solution exists.

\subsection{Langevin Equation}
We introduce stochastic integration via the Langevin equation.
In the presence of a viscous drag linearly proportional to velocity, the equation of motion for a particle of mass '$m$' is given by
\be \label{eq:newton}
 m\dot{v} + \alpha v = 0
\ee
or
\[ \dot{v} +\gamma v =0 \]
where $\gamma = \alpha/m = 1/\tau$, $\tau $ being the relaxation time.
 The solution of the above equation is given by
\[ v(t) = v(0) e^{-t/\tau} = v(0) e^{-\gamma t} \]

If the mass of the particle is small, so that the velocity due to thermal
fluctuations is not negligible, then
\[v_{th} = \sqrt{\langle v^2\rangle} = \sqrt{kT/m}\]
is observable and hence the velocity of the small paricle cannot be
described by Eqn. (\ref{eq:newton}). Thus this equation has to be modified as
follows:
\be \label{eq:langevin}
	\dot{v} +\gamma v = \Gamma(t)
\ee
where $\Gamma(t) = F_f(t)/m$ is the stochastic term,
 $F_f(t)$ is the fluctuating force acting on the particle. This is the Langevin equation.
\subsubsection{Brownian Motion}
The Langevin equation for Brownian motion is given by Eqn. (\ref{eq:langevin})
where $\Gamma$ describes the Langevin force with
\[ \langle \Gamma(t) \rangle = 0, \quad \langle \Gamma(t) \Gamma(t') \rangle = q \delta(t-t') \]
such that all the higher moments are given in terms of the two point
correlation function. In other words, $\Gamma$ is gaussian distributed with
zero mean. The spectral density of noise, described below, gives color of the
noise. The delta correlated noise is referred to as white noise. This model of
 noise in the Langevin equation fully describes ordinary Brownian Motion of a
 particle.

The spectral density for delta correlated noise is given by
\[S(\omega) = 2q \int_{-\infty}^\infty e^{-i\omega \tau} \delta({\tau}) d \tau
 = 2 q \]
Since it is independent of $\omega$ it is called white noise. In general the
spectral density depends on $w$; in such a case, the noise is called colored
noise.
\subsubsection{Solution of the Langevin equation}
The formal solution of the Langevin equation is given by
\be
v(t) = v_0 e^{-\gamma t} + \int_0^t e^{-\gamma(t-t')} \Gamma(t') dt'
\ee
By using the white noise model one can obtain the correlation function of
velocity
\begin{eqnarray}
& & \langle v(t_1)v(t-2)\rangle = v_0^2 e^{-\gamma(t_1+t_2)} +\nonumber \\ & & \int_0^{t_1} \int_0^{t_2}
e^{-\gamma(t_1+t_2-t_1'-t_2')} q \delta(t_1'-t_2') dt_1'dt_2' \nonumber \\
\end{eqnarray}
which after solving for the double integral gives
\begin{eqnarray}
& &\langle v(t_1)v(t_2) \rangle  = v_0^2 e^{-\gamma(t_1+t_2)} + \nonumber\\ & & \frac{q}{2 \gamma} (e^{-\gamma
|t_1-t_2|}- e^{-\gamma(t_1 +t_2)}).
\end{eqnarray}
For large $t_1$ and $t_2$, i.e $ \gamma t_1 , \gamma t_2 \gg 1$ the correlation
is independent of $v_0$ and is a function of the time difference.
\[\langle v(t_1) v(t_2)\rangle  = \frac{q}{2 \gamma} e^{-\gamma|t_1-t_2|} \]
Now in the stationary state
\[ \langle E \rangle = \frac{1}{2} m \langle v(t)^2 \rangle = \frac{1}{2} m \frac{q}{2 \gamma} \]
for a Brownian particle.

According to the law of equipartition
\[\langle E \rangle  = \frac{1}{2} kT \]
and comparing with the earlier expression we get
\[q = 2 \gamma kT /m\]
\subsubsection{Overdamped Langevin equation}
The overdamped Langevin Equation looks like
\[v(t) = \frac{1}{\gamma} \Gamma(t) \]
where the acceleration term is dropped because of the prominence of damping.
In the large time limit, the two point correlation for velocity is given by
\[\langle v(t_1) v(t_2)\rangle  \approx \frac{1}{\gamma^2} \langle \Gamma(t_1) \Gamma(t_2) \rangle =
\frac{q}{\gamma^2} \delta(t_1 - t_2) \]

\subsubsection{Non-linear Langevin Equation}
A nonlinear Langevin equation has the following form
\be \label{eq:nonlinear}
\dot{\xi}= h(\xi,t) + g(\xi,t)\Gamma(t)
\ee
Here $\Gamma(t)$ is assumed to be Gaussian random variable with zero mean and
$\delta$ correlation function.
\[\langle \Gamma(t)\rangle   = 0 ; \quad \langle \Gamma(t) \Gamma(t') \rangle = 2 \delta(t-t') \]

Integrating Eqn. (\ref{eq:nonlinear})
\be
\xi = \int_0^t h(\xi,t) dt + \int_0^t g(\xi,t) \Gamma(t) dt
\ee
Here '$g$' could be a constant or can depend on $\xi$. Constant $g$ gives
additive noise, while if it depends on $\xi$ it is referred to as multiplicative
 noise.
For $\xi$ dependent '$g$' in the above equation, since $\Gamma(t)$ has no
correlation time, it is not clear which value of $\xi$ one has to use in $g$
while evaluating the integral. Physicists use an approximation of the
 $\delta$ function to ensure that one gets appropriate results. But from a
purely mathematical point of view one cannot answer this question, unless some additional specification is given. Thus, this gives rise to the requirement of
 using the stochastic integrals, namely the It\^o and Stratonovich integrals
 for solving such equations. In the following section, we give a brief
introduction to these Stochastic Integrals. Prior to this we state an example of a particular form of noise and solve the nonlinear Langevin equation for this
 particular case.

\textbf{Example}
For $g = a \xi$ where $a$ is a constant
\[\dot{\xi} = a \xi \Gamma(t) \]
The formal solution of the above equation is given by
\[\xi(t) = x \exp\left[a \int_0^t \Gamma(t') dt' \right] \]
Assuming $ \xi(0) = x$
\begin{eqnarray}
& & \langle \xi(t) \rangle = \langle x \exp[a \int_0^t \Gamma(t') dt' ] \rangle \nonumber \\
& = & x[ 1+ a \int_0^t \langle \Gamma(t_1) \rangle dt_1 + \nonumber \\ &  &\frac{1}{2!} a^2 \int_0^t \int_0^t
\langle \Gamma(t_1) \Gamma(t_2) \rangle dt_1 dt_2 + \dots ]\nonumber\\
\end{eqnarray}

Since $\Gamma$ is delta correlated Gaussian white noise, all the higher
correlations can be expressed in terms of two point correlations, therefore the
 above equation can be written in the following form
\begin{eqnarray}
& = & \frac{(2n)!}{2^n n!} \left[ \int_0^t \int_0^t \phi(t_1 - t_2) dt_1 dt_2\right]^n \\
& = & \exp\left[\frac{1}{2} a^2 \int_0^t \int_0^t \langle \Gamma(t_1)\Gamma(t_2) \rangle dt_1
dt_2 \right] \nonumber \\
\end{eqnarray}
For delta correlated Langevin force the double integral gives

\[
\langle\xi(t) \rangle = x \exp(a^2 t)
\]
and
\[ \langle\dot{\xi}(t) \rangle  = a^2 \langle \xi(t)\rangle  \mbox{ with } \langle \xi(0)\rangle  = x \]
\[\frac{d}{dt} \langle  \xi(t)\rangle |_{t=0} = a^2 x \]  \mbox{ is called spurious or noise
induced drift. }

\subsection{Stochastic integration, It\^o calculus and stochastic differential equations}

Consider a Langevin equation of the form
\begin{equation}
\frac{dx}{dt} = a(x,t) + b(x,t) \zeta (t)
\end{equation}
for a time-dependent variable $x$, where $a(x,t)$ and $b(x,t)$ are known functions, and $\zeta (t)$ is the rapidly fluctuating random term which induces stochasticity in the evolution. We want to examine the mathematical status of this equation as a differential equation. Since we expect it to be integrable, the integral
\begin{equation}
u(t) = \int_{0}^{t} dt' \; \zeta (t')
\end{equation}
should exist. If we demand that $u$ is a continuous function of $t$, then it is a Markov process whose evolution can be described by a Fokker-Planck equation, which can be shown to have zero drift, and diffusion unity. Hence, $u(t)$ is a Wiener process, denoted say by $W(t)$; but we know that $W(t)$ is not differentiable. This would imply that in a mathematical sense the Langevin equation does not exist! However, the corresponding integral equation
\begin{equation}
x(t) - x(0) = \int_{0}^{t} a[x(s), s] ds + \int_{0}^{t} b[x(s), s] \zeta (s) ds
\end{equation}
can be interpreted consistently.  We make the replacement
\begin{equation}
dW(t) \equiv W(t+dt) - W(t) = \zeta (t) dt
\end{equation}
so that the second integral can be written as
\begin{equation}
\int_{0}^{t} b[x(s), s]\; dW(s)
\end{equation}
which is a stochastic Riemann-Stieltjes integral, which we now define.

Given that $G(t)$ is an arbitrary function of time and $W(t)$ is a stochastic process, the stochastic integral $\int_0^{t} G(t') \; dW(t')$ is defined by dividing the interval $(0,t)$ into $n$ sub-intervals $(t_{i-1},t_i)$ such that
\begin{equation}
t_0 \leq t_1 \leq t_2 \leq \dots \leq t
\end{equation}
and defining intermediate points $\tau_i$ such that
$t_{i-1} \leq \tau_i \leq t_i$. The stochastic integral is defined as a limit of partial sums
\begin{equation}
S_n = \sum_{i=1}^{n} G(\tau_i) [ W(t_i) - W(t_{i-1})].
\end{equation}
The challenge is that, $G(t)$ being function of a random variable, the limit of $S_n$ depends on the partcular choice of intermediate point $\tau_i$!  Different choices of the intermediate point give different results for the integral.

Denoting $\Delta= max(t_{i} - t_{i-1})$, the {\it It\^o stochastic integral } is defined by taking $\tau_{i}= t_{i-1}$ and  taking the limit of the sum :
\begin{eqnarray}
& & \int_{0}^{\tau} G(t') \; dW(t') = \nonumber \\ & & \lim_{\Delta\rightarrow 0}  \sum_{i=1}^{n} G(t_{i-1}) [ W(t_i) - W(t_{i-1})].
\end{eqnarray}
As an example, it can be shown that
\begin{eqnarray}
& & \int_{0}^{t} W(t') \; dW(t') = \nonumber \\ & &  \frac{1}{2} \left[ W(t)^2 - W(t_{0})^2  - (t-t_0)\right].
\end{eqnarray}
Note that the result for the integration is no longer the same as the ordinary Riemann-Stieltjes integral, where the term $(t-t_0)$ would be absent; the reason being that the difference $W(t + \Delta t) - W(t)$  is almost always of the order $\sqrt{t}$, which implies that, unlike in ordinary integration, terms of second order in $\Delta W(t)$ do not vanish on taking the limit.

An alternative definition of the stochastic integral is the {\it Stratanovich integral}, denoted by $S$, and is such that the anomalous term above, $(t-t_{0})$, does not occur. This happens if the intermediate point $\tau_i$ is taken as the mid-point $\tau_{i} = (t_i + t_{i+1}))/2$, and it can then be shown that
\begin{equation}
S\int_{0}^{t} W(t') dW(t') = \frac{1}{2} [ W(t)^2 - W(t_0)^2 ].
\end{equation}

For arbitrary functions $G(t)$ there is no connection between the It\^o integral and the Stratanovich integral. However, in cases where we can specify that $G(t)$ is related to some stochastic differential equation, a formula can be given relating the two differential equations.

\subsubsection{Rules of It\^o calculus}

Mean square limit or the limit in the mean is defined as follows:
Let $X_n(\omega)$ be a sequence of random variables $X_n$ on the probability space $\Omega$, where $\omega$ are the elements of the space which have  probability density  $p(\omega)$.  Thus one can say that $X_n$ converges to $X$ in the
\textit{mean square} if
\begin{eqnarray}
& & \lim_{n\rightarrow \infty} \int d\omega p(\omega) [X_n(\omega) -X(\omega)]^2
\equiv \nonumber\\ & & \lim \langle (X_n - X)^2\rangle = 0
\end{eqnarray}
This is written as
\[ ms-\lim_{n \rightarrow \infty} X_n = X \]
\textbf{Rules}
\begin{enumerate}
\item
$ dW(t)^2 = dt$
\item
$ dW^{2+N}(t) =0 $
\end{enumerate}
The above formulae mean the following
\begin{eqnarray}
& & \int_{t_0}^t [dW(t')]^{2+N} G(t')   \equiv  \nonumber \\ & & ms-\lim{n \rightarrow \infty} \sum_i
G_{i-1} \Delta W_i^{2+N} \nonumber \\
&=& \int_{t_0}^t dt' G(t') \mbox{ for }  N =0 \nonumber \\
& = & 0 \mbox{ for } N > 0
\end{eqnarray}
for an arbitrary non-anticipating function $G$.

\textbf{Proof:}

For $N =0$ consider the following sum
\begin{eqnarray}
& & I  =  \lim_{n \rightarrow \infty} \langle[\sum_i G_{i-1} (\Delta W_i^2 -
 \Delta t_i)]^2] \rangle\\
& & =  \lim_{n \rightarrow \infty}  \langle \sum_i (G_{i-1})^2(\Delta W_i^2 - \Delta
t_i)^2 + \nonumber \\ & & \sum_{i > j} 2 G_{i-1} G_{j-1} (\Delta W_j^2 - \Delta t_j)(\Delta W_i^2 - \Delta t_i)\rangle  \nonumber \\
\end{eqnarray}

Using the following results mentioned earlier
\[ \langle\Delta W_i^2\rangle = \Delta t_i \]
\[ (\Delta W_i^2- \Delta t_i)^2 > = 2 \Delta t_i^2 \]
one gets
\[ I  = 2 \lim_{ n \rightarrow \infty} [\sum_i \Delta t_i^2 \langle (G_{i-1})^2\rangle] \]
This can be written as
\[ ms -\lim_{n \rightarrow \infty} (\sum_i G_{i-1} \Delta W_i^2 - \sum_i
G_{i-1} \Delta t_i) =0 \]
since
\[ms-\lim_{n \rightarrow \infty} \sum_i G_{i-1} \Delta t_i = \int_{t_0}^t
dt' G(t') \]
we get
\[\int_{t_0}^t [dW(t')]^2 G(t') = \int_{t_0}^t dt' G(t')\]
from this we see that
$ dW(t')^2 = dt' $. Similarly one can show that $dW(t)^{2+N} \equiv 0 (N >0)$.

Another important result that can be proved by the above method is
\begin{eqnarray}
& & \int_{t_0}^t G(t') dt' dW(t') \equiv \nonumber \\ & & ms-\lim_{n \rightarrow \infty} \sum
G_{i-1} \Delta W_i \Delta t_i = 0
\end{eqnarray}

Rule for integration of polynomials :
\begin{eqnarray}
& & \int_{t_{0}}^{t} W(t')^n \; dW(t') \nonumber \\ = & & \frac{1}{n+1} [W(t)^{(n+1)} - W(t_0)^{(n+1)}]\nonumber \\
& & - \frac{n}{2} \int t_{0}^{t} W(t)^(n-1)\; dt
\end{eqnarray}

General rule for differentiation ;
\begin{eqnarray}
& & df[W(t), t] = \left( \frac{\partial f}{\partial t} + \frac{1}{2}\frac{\partial ^ 2 f}{\partial W^2} \right) dt + \nonumber \\ & & \frac{\partial f}{\partial W} dW(t).
\end{eqnarray}

\subsubsection{Stochastic differential equations}
The It\^o integral is mathematically the most satisfactory, but not always the most natural physical choice. The Stratanovich integral is the natural choice for an interpretation where $\zeta(t)$ is a colored (not white) noise. Also, unlike in the It\^o interpretation, the Stratanovich interpretation enables the use of ordinary calculus. From the mathematical point of view, it is more convenient to define the It\^o SDE, develop its equvalence with the Stratanovich SDE, and use either form depending on circumstances.

A stochastic quantity $x(t)$ obeys an It\^o differential equation
\begin{equation}
dx(t) = a[x(t), t] dt + b[x(t), t]dW(t)
\label{itosde}
\end{equation}
if for all $t$ and $t_0$
\begin{eqnarray}
& & x(t) = x(t_0) + \int _{t_{0}} ^{t} dt ' a[x(t'), t'] + \nonumber \\ & & \int _ {t_{0}}^{t} dW(t') b[x(t'), t'].
\end{eqnarray}
If $f[x(t)]$ is an arbitrary function of $x(t)$ then It\^o's formula gives the differential equation satisfied by f :
\begin{eqnarray}
& & df[x(t)] = \nonumber \\ & & \left(a[x(t),t] f'[x(t)] + \frac{1}{2} b[x(t), t]^2 f"[x(t)]\right) dt \nonumber \\ &  & +
b[x(t), t] f'[x(t)]dW(t).
\end{eqnarray}
Thus change of variables is not given by ordinary calculus unless $f[x(t)]$ is linear in $x(t)$.

Given the time development of an arbitrary $f[x(t)]$, the conditional probability density $p(x,t|x_0, t_0)$ for $x(t)$ can be shown to satisfy a Fokker-Planck equation with drift coefficient $a(x,t)$ and diffusion coefficient $b(x,t)^2$.

The stochastic differential equation studied in detail in the QMUPL model ~\cite{RMP:2012}
\begin{eqnarray} \label{eq:qmupl1repeat}
d \psi_t & = &  \left[ -\frac{i}{\hbar} H dt + \sqrt{\lambda} (q - \langle q \rangle_t) dW_t \right. \nonumber \\
& & - \left. \frac{\lambda}{2} (q - \langle q \rangle_t)^2 dt \right] \psi_t,
\end{eqnarray}
is an It\^o differential equation of the type (\ref{itosde}). If we formally treat this as an equation for $\ln\psi$ then upon comparison we see that $\ln\psi$ is $x$, and
\begin{eqnarray}
& & a(x(t),t] = -\frac{i}{\hbar} H - \frac {\lambda}{2} (q - \langle q \rangle_t) ^2  \nonumber \\
& & b = \sqrt{\lambda} (q - \langle q\rangle)_t
\end{eqnarray}
The real part of the drift coefficient $a$ is related to the diffusion coefficient $b^2$ by
\begin{equation}
-2a = b^2
\end{equation}
This non-trivial relation between diffusion and drift is what gives the equation its norm-preserving martingale property which eventually leads to the Born rule. Why the drift and diffusion must be related this way is at present not understood in Trace Dynamics and there probably is some deep underlying reason for this relation.

Stratonovich's stochastic differential equation : Given the It\^o differential equation (\ref{itosde}) its solution $x(t)$ can also be expressed in terms of a Stratonovich integral
\begin{eqnarray}
& & x(t) = x_{0} + \int_{t_{0}}^{t} dt' \alpha [x(t'), t'] + \nonumber \\ &  & S\int _{t_{0}}^{t} dW(t') \beta[(x(t'),t']
\end{eqnarray}
where
\begin{equation}
\alpha(x,t) = a(x,t) - \frac{1}{2} b(x,t)\partial_{x} b(x,t);   \beta(x,t) = b(x,t)
\end{equation}
 In other words, the It\^o SDE (\ref{itosde}) is the same as the Stratonovich SDE
\begin{equation}
dx = [a(x,t) - \frac{1}{2} b(x,t)\partial_{x} b(x,t)] dt + b dW (t)
\end{equation}
or conversely, the Stratonovich SDE
\begin{equation}
dx = \alpha dt + \beta dW(t)
\end{equation}
is the same as the It\^o SDE
\begin{equation}
dx = [\alpha(x,t) + \frac{1}{2} \beta (x,t)\partial_{x} \beta (x,t)] dt + \beta dW (t)
\end{equation}
It can be shown that in a Stratonovich SDE the rule for change of variable is the same as in ordinary calculus.

\subsection{Martingales}
Martingales play a role in stochastic processes roughly similar to
that played by conserved quantities in dynamical systems. Unlike a
conserved quantity in dynamics, which remains constant in time, a
martingale's value can change ; however, its expectation remains
constant in time.

A martingale is defined as follows:

A discrete time martingale is a discrete time stochastic process, $X_1
,X-2, \dots , $ that satisfies for any time $n$,
\begin{eqnarray*}
E(|X_n|)& < & \infty \\
E(X_{n+1}|X_1, \dots , X_n ) &  = &  X_n
\end{eqnarray*}
where $E( X) $ denotes the expectation of $X$.

\textbf{Martingale Sequence with Respect to Another Sequence}
A sequence $Y_1,Y_2, \dots$ is said to be a martingale w.r.t. another
sequence $X_1,X_2,X_3 \dots $ if for all $n$
\begin{eqnarray*}
E(|Y_n|) & < &  \infty \\
E(Y_{n+1} |X_1,\dots , X_n) & = & 0
\end{eqnarray*}
In other words, a martingale is a model of a fair game, where no
knowledge of past events can help to predict future winnings. It is a sequence
of random variables for which at a particular time in a realized sequence, the
expectation of the next value in the sequence is equal to the present
observed value even given knowledge of all prior observed value at current
time.

\bibliography{biblioqmts3}

\end{document}